\documentstyle[prl,aps]{revtex}
\begin{document}
\draft

\twocolumn[\hsize\textwidth\columnwidth\hsize\csname@twocolumnfalse\endcsname

\title{Extremal dynamics model on evolving networks}

\author{Franti\v{s}ek Slanina \cite{f-adr}}
\address{Institute of Physics, Academy of Sciences of the Czech Republic,\\
Na Slovance 2, CZ-182~21 Praha 8, Czech Republic\\
and Center for Theoretical Study, Jilsk\'a 1, CZ-11000 Praha 1
, Czech Republic}

\author{Miroslav Kotrla \cite{m-adr}}
\address{Institute of Physics, Academy of Sciences of the Czech Republic,\\
Na Slovance 2, CZ-182~21 Praha 8, Czech Republic}

\maketitle
\begin{abstract}
We investigate an extremal dynamics model of 
evolution
with  variable number of units.
Due to addition and removal of the units, the topology of the network
evolves and the network splits into several clusters. 
The activity is
mostly concentrated in the  largest cluster.
The time dependence
of the number of units exhibits intermittent structure.
The self-organized criticality is manifested by a power-law distribution
of forward avalanches, but two regimes with distinct exponents
$\tau=1.98\pm 0.04$ and $\tau^\prime = 1.65\pm 0.05$ are found.
The distribution of
extinction sizes obeys a power law with exponent $ 2.32\pm 0.05$. 
 \end{abstract}
\pacs{PACS numbers: 05.40.-a, 87.10.+e, 87.23.Kg}

\twocolumn]

Extremal dynamics (ED) models \cite{pa_ma_ba_96} are used in wide area
of problems, ranging 
from growth in disordered medium \cite{sneppen_92}, dislocation
movement \cite{za_92}, friction \cite{slanina_98} to biological
evolution \cite{ba_sne_93}. Among them, the Bak-Sneppen (BS) model
\cite{ba_sne_93} plays the role of a testing ground for various
analytical as well as numerical approaches (see for example
\cite{pa_ma_ba_96,grassberger_95,bo_ja_we_95,pismak_95,ma_lor_ma_97,%
delos_va_ve_97,va_au_95}).
The 
dynamical system in question is composed of a large number of
simple units, connected in a network. Each site of the network hosts
one unit. The state of each unit is described
by a dynamical variable $b$, called barrier. In each step,
the unit with minimum $b$ is mutated by updating the barrier.
The effect of the mutation on the
environment is taken into account by changing $b$ also at all
neighbors of the minimum site.

General feature of
ED models is the avalanche dynamics. The forward $\lambda$-avalanches
are defined as follows \cite{pa_ma_ba_96}. For fixed $\lambda$ we define
active sites as those having barrier $b < \lambda$. Appearance
of one active site can lead to avalanche-like proliferation of active
sites in successive time steps. The avalanche stops, when all active
sites disappear  
again. There is a value of $\lambda$, for which the probability
distribution of  avalanche sizes obeys a power law without any
parameter tuning, so  that the ED models are classified as a subgroup
of self-organized critical models \cite{ba_ta_wi_87}. 

The BS model was originally devised in order to explain the
intermittent structure of the extinction events seen in the fossil
record \cite{sneppen_95}. In various versions of the BS model it was
found, that the 
avalanche exponent is $1<\tau \le 3/2$, where the maximum value $3/2$
holds in the mean-field universality class. On the other hand, in
experimental data for the distribution of extinction sizes higher
values of the exponent, typically around 
$\tau\simeq 2$ are found
\cite{newman_97a}. The avalanche exponent close to 2 was also measured
in ricepile experiments \cite{fre_chri_ma_fe_jo_mea_96}.
 While there are several
models of different kind, which give generic value $\tau=2$
\cite{ne_sne_96,so_ma_96}, we 
are not aware of 
any ED  model with such a big value of the exponent.

The universality class a particular model belongs
to, depends on the topology of the network on which the units
are located. Usually, regular hypercubic networks \cite{pa_ma_ba_96}
or Cayley trees 
\cite{va_au_95} are 
investigated. For random neighbor networks, mean-field solution
was found to be exact
\cite{debo_de_fly_ja_we_94,bo_ja_we_95}. Also the tree models  \cite{va_au_95}
were found to belong to the mean-field universality class.
Recently, BS model on random networks, produced by bond percolation on
fully connected lattice, was studied \cite{chri_do_ko_sne_98}. Two
universality classes were found. Above the percolation threshold, the
system belongs to the mean-field
universality class, while exactly at the
percolation threshold, the avalanche exponent is different.
 A dynamics changing the topology in order to drive the
network to critical connectivity was suggested. Similar model was
investigated recently \cite{ja_kri_98} in the context of autocatalytic
chemical reactions.

We present here a further step towards reality.
In fact, one can find real systems in which not only the topology of
connections 
evolves, but also the number of units changes. The network develops
new connections, when a new unit is inserted, and if a unit is removed,
its links are broken. This is a typical situation in natural ecologies,
where each extinction and speciation event changes also the 
topology of the ecological network. The same may apply to economics
and other areas, where the range of interaction is not determined by
physical Euclidean space. 
This 
problem was already 
partially investigated within mean-field BS model 
\cite{he_ro_97} and also in several other models devised for
description of biological evolution
\cite{so_ma_96,wi_ma_97,abramson_97,am_me_98}, which use
different approaches than the extremal dynamics. 

The purpose of this Letter is twofold. First, to study the evolution
of topology in a
ED model with variable number of units. Second, to demonstrate that the
large value of the avalanche exponent can be observed if the  topology
of the underlying network evolves dynamically.

We consider a system composed of varying number $n_{\rm u}$ of
units connected in a network. In the context of biological evolution,
these units are species.  
The dynamical rules of our model are the following.

{\it (i) } Each unit has a barrier $b$ against mutations. The unit
with minimum $b$ is mutated.

{\it (ii) } The barrier of the mutated unit is replaced by a new random
value $b^\prime$. Also the barriers of all its neighbors are replaced
by new random numbers. 
If $b^\prime$ is larger than barriers of all its neighbors,
the unit gives birth to a new unit (speciation). If $b^\prime$ 
is lower than barriers of all neighbors, the 
unit dies out (extinction). As a boundary condition, we use the
following exception: 
if the network
consists of a single isolated unit only, it never dies out.

This rule is motivated by the following considerations. We assume,
that well-adapted units 
proliferate more rapidly and chance for speciation is bigger. However,
if the local biodiversity, measured by connectivity of the unit, is
bigger, there are fewer empty ecological niches and the probability of 
speciation is lower. On the other hand,
poorly adapted units are more vulnerable to extinction, but at the
same time larger biodiversity (larger connectivity) may favor the survival.
Our rule corresponds well to these assumptions: speciation occurs
preferably at units with high barrier and surrounded by fewer
neighbors, 
extinction is more frequent at units with lower barrier and lower
connectivity. Moreover, we suppose that a unit completely isolated
from the rest of the ecosystem has very low chance to survive. This
leads to the following rule.

{\it (iii) } If a unit dies out, all its neighbors which
are not connected to any other unit also die out. We call this kind of
extinctions singular extinctions.

From the rule {\it (ii)} alone follows equal probability of
adding and removing a unit, while the rule {\it (iii)} enhances the
probabilty of the removal.
As a result, the probability of speciation is slightly lower than
the probability of extinction. 
The degree of disequilibrium between
the two
depends on the topology of the network at the moment and can be
quantified by the frequency of singular extinctions. The number of
units $n_{\rm u}$ perform a biased random walk with reflecting
boundary at $n_{\rm u} = 1$. The bias towards small values is not
constant, though, but fluctuates as well. 

{\it (iv) } Extinction means, that the unit is removed
without 
any substitution and all links it has, are broken.
Speciation means, that a new unit is added
into the system, with a random barrier. The new unit is connected to
all neighbors of the mutated unit:  all links of the
``mother'' unit are inherited by the ``daughter'' unit.
This rule reflects the fact that the new unit is to a certain extent 
a copy of the 
original, so the relations to the environment will be initially
similar to the ones the old unit has. 
Moreover,
if a unit which speciates has only one neighbor, a link between
``mother'' and ``daughter'' is also established.

\begin{figure}[hb]
  \centering
  \vspace*{60mm}
  \includegraphics{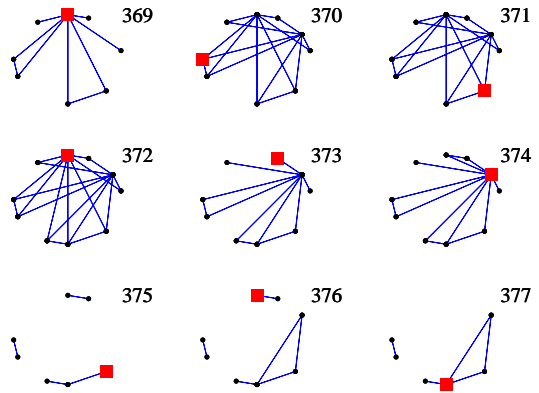}
  \caption{An example of network configurations in several successive
  time steps, from step 369 to 377. Units are represented by small
  filled circles. The lines represent links between the units. The unit
  with the minimal barrier $b$ is indicated by a filled square.}
  \label{fig:sequence}
\end{figure}

The above described rules are illustrated in
Fig. \ref{fig:sequence}. Speciation occurs in the transition from step
369 to 370, 371 to 372, 373 to 374 and 375 to 376. In the transitions
from 372 to 373, 374 to 375 and 376 to 377 extinction occurs (last two
include also singular extinctions). We can see that after speciation
the neighbors of the new unit have one neighbor more than 
before, so that if $n_{\rm u}$ increases, also the connectivity
of the network grows.

\begin{figure}[hb]
  \centering
  \vspace*{100mm}
  \includegraphics{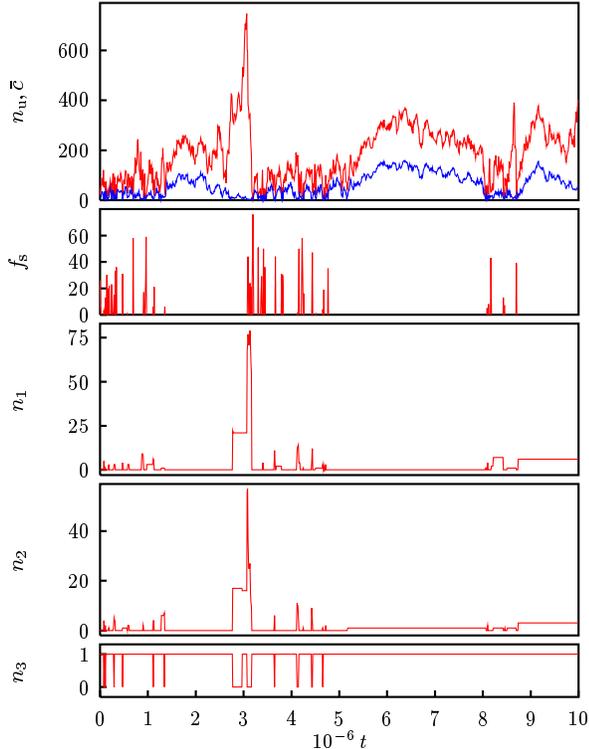}
  \caption{Time evolution of quantities describing the size and
  topology of the network, in a typical run. From top to bottom: the
  number of units   $n_{\rm u}$
  (upper line) and average connectivity $\bar{c}$ (lower line), 
  number of singular
  extinctions during $10^3$ steps, number of clusters of size
  2, number of clusters of size larger than 2 and  $\le n_{\rm
  u} /2$, number of clusters of size larger than $n_{\rm
  u} /2$.}
  \label{fig:num-and-other}
\end{figure}

We investigated the evolution of the network by measuring time
dependence of several quantities. We start the simulation with the
initial condition $n_u=1$. A typical result is shown in 
Fig. \ref{fig:num-and-other}, where we show the time dependence of 
the number of units $n_{\rm u}$, average connectivity $\bar{c}$,
and the
frequency of singular extinctions $f_{\rm s}$. 
The network can be split into
disconnected clusters, as is illustrated in the Fig. \ref{fig:sequence}. 
In Fig. \ref{fig:num-and-other} we show also
the evolution of number of
clusters in three size categories.  We denote by $n_1$ the number of 
the smallest clusters, of size 2, by $n_2$ the number of
medium-size clusters, larger than 2 and smaller or equal
to $n_{\rm u} /2$, and by $n_3$ the number of
clusters larger than $n_{\rm u} /2$. The value $n_3=1$ means that most
of the system is concentrated in a single cluster. 

We can see that
singular extinctions occur in bursts. The passages without singular
extinctions, where number of units evolves like a random walk (due to
equal probability of increase and decrease of number of units by 1)
are interrupted by short periods, where $n_{\rm u}$ falls to small
values and singular extinctions are intense. 
We can see that very often three events coincide: high frequency of
singular extinctions, large number of clusters, especially of size 2,
and the fact, that the largest cluster does not contain most of the network.
We observed, that the mutation occurs  nearly
all the time in the largest cluster.
A similar effect was reported also in the Cayley tree models
\cite{va_au_95}: the small isolated portions of the network are
very stable and nearly untouched by the evolution.

\begin{figure}[hb]
  \centering
  \vspace*{50mm}
  \includegraphics{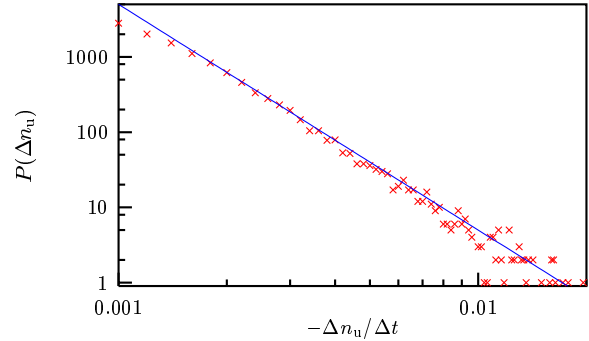}
  \caption{Distribution of drops in number of units
  during the interval $\Delta t=3\cdot 10^4$ steps. The full line is a
  power with exponent -3.}
  \label{fig:hist}
\end{figure}

Figure \ref{fig:num-and-other} suggests that the number of units
exhibits intermittent drops. This corresponds
qualitatively to punctuated  
equilibria seen in the fossil data. This feature is new in our model,
when compared with previous approaches within the BS model
\cite{he_ro_97} but resembles the intermittent features of models
based on neural networks \cite{so_ma_96}, Lotka-Volterra equations
\cite{abramson_97}, and 
coherent noise \cite{wi_ma_97}.  In order to check this
property quantitatively, we plot in Fig. \ref{fig:hist} the
distribution of changes in number of units during time
interval $\Delta t = 3\cdot 10^4$ steps. 
We can see that the distribution of drops ($\Delta n_{\rm u}<0$) has a
power-law tail, which confirms the intermittency.

\begin{figure}[hb]
  \centering
  \vspace*{47mm}
  \includegraphics{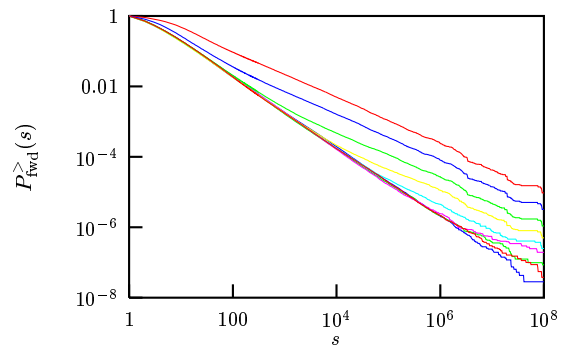}
  \caption{Distribution of forward avalanche sizes, averaged over 12
  independent runs.
  Each run lasted $3\cdot 10^8$ steps. The values 
  of $\lambda$ are (from bottom to top)
  0.013, 0.016, 0.02, 0.03, 0.05, 0.1, 0.2, 0.4, 0.6. The
  superscript $>$ indicates that we count all events larger than $s$.}
  \label{fig:aval}
\end{figure}

The distribution of forward $\lambda$-avalanches
\cite{pa_ma_ba_96} is shown in Fig. \ref{fig:aval}. We found 
power-law distribution for  
$\lambda_c= 0.016$ with the exponent $\tau= 1.98\pm 0.04$. 
The value of the exponent was found by fitting the data in the interval
$(500,5\cdot 10^5)$\cite{note:errorbars}.  
Contrary to the BS and related models, we found power law
distribution with a different exponent $\tau^\prime=1.65\pm 0.05$  
also for $\lambda$ larger than about 0.4. (More precisely, we obtained
the value of the exponent by fitting the data in the interval
$(500,10^6)$ for $\lambda=0.4$ and $\lambda=0.6$. Both values
of $\lambda$ give the same result).  
 The data suggest that for $\lambda >\lambda_c$ the 
avalanche size distribution  exhibits two regimes, with crossover
around certain avalanche size $s_{\rm cross}$. For small
avalanches, $s<s_{\rm cross}$, the distribution is power law with
exponent $\tau$, while for larger avalanches, $s>s_{\rm cross}$, power
law with exponent $\tau^\prime$ holds. The crossover $s_{\rm cross}$
grows when $\lambda$ approaches to $\lambda_c$.

The existence of avalanches for $\lambda$ close to 1 is
related to the fluctuation of the number of units. We observed, that
such avalanches start and end mostly when number of units is very
small. Between these events the evolution of the number of units is
essentially a random walk, because singular extinctions are rare. This fact
can explain, why the exponent $\tau^\prime$ is not too far from the
value $3/2$ corresponding to the distribution of first returns to the
origin for the random walk. The difference is probably due to the
presence of singular extinctions.

The fact, that the number of units change, enables us to define the
extinction size in a more realistic way than in the previous variants
of the BS model.
For fixed $\lambda$ we count number of units
which were present at the beginning of the $\lambda$-avalanche and are
no more 
present when the avalanche stops. This quantity corresponds better to
the term ``extinction size'' used by paleontologists than the number
of units affected by mutations, as it is defined in the BS model. 

\begin{figure}[hb]
  \centering
  \vspace*{47mm}
  \includegraphics{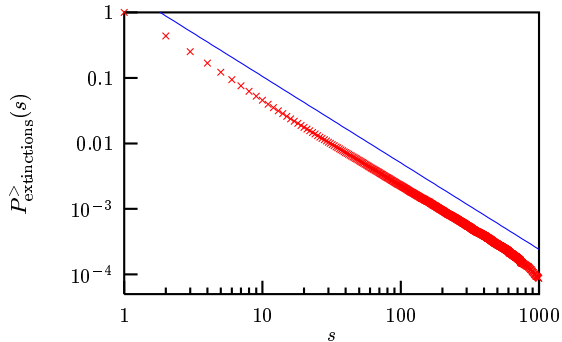}
  \caption{Distribution of extinction sizes for $\lambda = 0.016$,
  averaged over 12
  independent runs. Each run lasted
  $3\cdot 10^8$ steps.
  The full line corresponds to exponent $\tau_{\rm ext}=2.32$. The
  superscript $>$ indicates that we count all events larger than $s$.}  
  \label{fig:extinctions}
\end{figure}

For $\lambda=\lambda_c=0.016$ the distribution of extinction
sizes follows the power law with exponent 
$\tau_{\rm ext}=2.32\pm 0.05$, as is shown in Fig. \ref{fig:extinctions}.
 (The value of the
exponent was obtained by fitting the data in the interval $(10,1000)$.) 
This value is larger than the exponent 2 observed in the statistics of
real biological extinctions, but still closer than the values found in
previous modifications of the BS model.

The fact,
that the network evolves and the number of units fluctuates
leads to significantly larger values of the exponent than in the BS
model, even greater that the 
experimental one, while variants of the BS model have values lower than
the experimental 
one. This suggests, that the freedom in changing the topology in our
model is exaggerated and in order to obtain a more realistic model of
the biological evolution, we should look for some principles, which
imply freezing of the topology, while allowing the species to be
replaced by new 
ones.

We studied several other modifications of our
model, in order to check its robustness. For example, 
the link between
``mother'' and ``daughter'' unit was established, 
or only certain fraction of the links connecting ``mother'' to its
neighbors was inherited by ``daughter''. These modifications affected
some aspects of the network dynamics, but the avalanche and extinction
statistics was not significantly different.

To sum up, we formulated and studied the extremal dynamics model
derived from the Bak-Sneppen model,
which exhibits 
forward-avalanche exponent close to two, due to the annealed topology
of the network. The extinction size was defined in more
realistic manner compared to previous approaches within the BS model
and the extinction statistics was found to obey a power law with
exponent somewhat larger than two. The value found is closer to
paleontological data than in the previous variants of the BS model.

We wish to thank A. Marko\v{s} for useful discussions. 







\begin{thebibliography}{99}
\bibitem{note:errorbars}
To establish the error bars $\delta$ for the exponent $\alpha$ (which
can be $\tau$, $\tau^\prime$ or $\tau_{\rm ext}$) we used the
following procedure.  
Let $I=(a,b)$ be the interval, in which $\alpha$ was determined by
fitting the data. 
We choose a moving subinterval $I^\prime=(a^\prime,b^\prime)\subset I$
with fixed ratio $b^\prime/a^\prime$
and find the value $\alpha^\prime$ of the same exponent by fitting in
$I^\prime$.  
The error bars are then established by the requirement that 
$\alpha^\prime\in (\alpha-\delta,\alpha+\delta) $ for any choice of
the subinterval $I^\prime$.
For the exponents $\tau,\tau^\prime$, we used
$b^\prime/a^\prime=10$, so that $I^\prime$ covers one
decade, while for $\tau_{\rm ext}$ we set $b^\prime/a^\prime=\sqrt{10}$.
\end{thebibliography}
\end{document}